\documentclass[]{article}
\usepackage{lmodern}
\usepackage{amssymb,amsmath}
\usepackage{ifxetex,ifluatex}
\usepackage{fixltx2e} 
\ifnum 0\ifxetex 1\fi\ifluatex 1\fi=0 
  \usepackage[T1]{fontenc}
  \usepackage[utf8]{inputenc}
\else 
  \ifxetex
    \usepackage{mathspec}
  \else
    \usepackage{fontspec}
  \fi
  \defaultfontfeatures{Ligatures=TeX,Scale=MatchLowercase}
\fi
\IfFileExists{upquote.sty}{\usepackage{upquote}}{}
\IfFileExists{microtype.sty}{%
\usepackage{microtype}
\UseMicrotypeSet[protrusion]{basicmath} 
}{}
\usepackage[margin=1in]{geometry}
\usepackage{hyperref}
\hypersetup{unicode=true,
            pdftitle={Social network aided plagiarism detection},
            pdfauthor={Aljaž Zrnec and Dejan Lavbič},
            pdfborder={0 0 0},
            breaklinks=true}
\usepackage{natbib}
\usepackage{longtable,booktabs}
\usepackage{graphicx,grffile}
\makeatletter
\def\maxwidth{\ifdim\Gin@nat@width>\linewidth\linewidth\else\Gin@nat@width\fi}
\def\maxheight{\ifdim\Gin@nat@height>\textheight\textheight\else\Gin@nat@height\fi}
\makeatother
\setkeys{Gin}{width=\maxwidth,height=\maxheight,keepaspectratio}
\IfFileExists{parskip.sty}{%
\usepackage{parskip}
}{
\setlength{\parindent}{0pt}
\setlength{\parskip}{6pt plus 2pt minus 1pt}
}
\providecommand{\tightlist}{%
  \setlength{\itemsep}{0pt}\setlength{\parskip}{0pt}}
\ifx\paragraph\undefined\else
\let\oldparagraph\paragraph
\renewcommand{\paragraph}[1]{\oldparagraph{#1}\mbox{}}
\fi
\ifx\subparagraph\undefined\else
\let\oldsubparagraph\subparagraph
\renewcommand{\subparagraph}[1]{\oldsubparagraph{#1}\mbox{}}
\fi

\let\rmarkdownfootnote\footnote%
\def\footnote{\protect\rmarkdownfootnote}

\usepackage{titling}

  \title{Social network aided plagiarism detection}
    \pretitle{\vspace{\droptitle}\centering\huge}
  \posttitle{\par}
    \author{Aljaž Zrnec and Dejan Lavbič}
    \preauthor{\centering\large\emph}
  \postauthor{\par}
    \date{}
    \predate{}\postdate{}

\usepackage{booktabs}
\usepackage{longtable}
\usepackage{array}
\usepackage{multirow}
\usepackage[table]{xcolor}
\usepackage{wrapfig}
\usepackage{float}
\usepackage{colortbl}
\usepackage{pdflscape}
\usepackage{tabu}
\usepackage{threeparttable}
\usepackage{threeparttablex}
\usepackage[normalem]{ulem}
\usepackage{makecell}

\usepackage{amsthm}

\theoremstyle{definition}

\theoremstyle{definition}

\theoremstyle{definition}

\theoremstyle{remark}

\begin{document}
\maketitle

\begin{quote}
Aljaž Zrnec and \textbf{Dejan Lavbič}. 2018.
\href{https://doi.org/10.1111/bjet.12345}{\textbf{Social network aided
plagiarism detection}},
\href{https://onlinelibrary.wiley.com/journal/14678535}{British Journal
of Educational Technology \textbf{(BJET)}}, 48, pp.~113 - 128.
\end{quote}

\section*{Abstract}\label{abstract}
\addcontentsline{toc}{section}{Abstract}

The prevalence of different kinds of electronic devices and the volume
of content on the Web have increased the amount of plagiarism, which is
considered an unethical act. If we want to be efficient in the detection
and prevention of these acts, we have to improve today's methods of
discovering plagiarism. The paper presents a research study where a
framework for the improved detection of plagiarism is proposed. The
framework focuses on the integration of social network information,
information from the Web, and an advanced semantically enriched
visualization of information about authors and documents that enables
the exploration of obtained data by seeking of advanced patterns of
plagiarism. To support the proposed framework, a special software tool
was also developed. The statistical evaluation confirmed that the
employment of social network analysis and advanced visualization
techniques led to improvements in the confirmation and investigation
stages of the plagiarism detection process, thereby enhancing the
overall efficiency of the plagiarism detection process.

\section*{Keywords}\label{keywords}
\addcontentsline{toc}{section}{Keywords}

Social Network Analysis, Plagiarism detection process, Plagiarism
visualization, Assessment, Facilitation, Learning management systems,
Student-centredness, Case study

\section{Introduction}\label{introduction}

The widespread availability of computers and mobile devices and the
volume of content on the Web have changed the approaches to both
teaching and learning processes. Simultaneously, the amount of
plagiarism has increased enormously over the last few years due to the
aforementioned changes. The act of plagiarism is defined as the
unethical action of copying someone else's work
\citep{youmans_does_2011}, and is usually considered as an offense,
therefore we also have to improve the current plagiarism detection
processes to be able to cope with the increasing amount of plagiarism in
a more efficient way.

Culwin and Lancaster defined the Four-Stage Plagiarism Detection Process
(FSPDP) \citep{culwin_visualising_2001} used to systematically search
for plagiarisms in a given set of documents focusing not only on
similarity detection. FSPDP consists of four stages: collection,
analysis, confirmation and investigation. Usually, all stages were
performed by a human investigator, but with the advent of different
plagiarism detection methods supported by computers, the first two
stages in this process can be fully automated, and the latter two can
only be partly automated \citep{makuc_methods_2013}. The effectiveness
of detection depends not only on the similarity engine
\citep{ali_overview_2011, hage_comparison_2010} used in the second
stage, but also on the rate of automation of the latter two stages in
the process. Any similarity in the second stage, which is considered as
positive, is further reviewed in the investigation stage. For a
submission to be judged to contain plagiarism, the confirmation stage
(3rd stage) must be completed, where the submission is examined and
verified by a human investigator. This stage can also be fully
automated, but false positive and false negative results may occur.

Today, most of the approaches \citep{ali_overview_2011} to the detection
of plagiarism are focused on the first two stages, namely collection and
analysis, leaving the investigator to perform the latter stages
manually. That was the motivation behind conducting research work to
propose a novel approach focusing on social aspects of potential
plagiarists, by taking into account their social network connections,
activities and information from the Web, to support investigator's work
in the third and the fourth stages of FSPDP, thereby making the
plagiarism detection process more efficient. We believe that the
plagiarism detection process can be improved by reducing the number of
manual examinations of potentially plagiarized work. This could be
achieved by the employment of new visualization techniques that enable a
semantically enriched view of the relationships between possible
plagiarists.

The paper is organized as follows. In the second section, we present the
related work and propose a solution. In the third section, we describe
our Social Plagiarism Detection Framework and supportive software tool.
In the fourth section, we present the evaluation method for assessing
the approach and discuss the obtained results. Finally, in the last
section, we conclude the paper and discuss the possibilities for future
research.

\section{Related work}\label{related-work}

\subsection{Review of related approaches and
tools}\label{review-of-related-approaches-and-tools}

According to authors \citep{mozgovoy_automatic_2010}, there are five
different types of plagiarism, varying from verbatim copying to advanced
types of plagiarism \citep{witherspoon_undergraduates_2012} such as the
copying of ideas and plagiarism in the form of translated text. The
increasing use of computers and Web 2.0 tools have mainly had a positive
effect on learning, but they also increase the possibility of using
different types of plagiarism \citep{underwood_academic_2003}.

With the expansion of various types of plagiarism, especially with the
proliferation of digital documents on the Web and the advent of social
networks, many innovative approaches to plagiarism detection have also
emerged. Several successful studies have been applied to traditional
approaches
\citep{mozgovoy_automatic_2010, mozgovoy_fast_2005, stein_intrinsic_2011},
focusing on program code or plain text.

Early approaches to plagiarism detection heavily relied on methods that
were based on string matching. Advanced methods include document parsing
to extract the structure of the sentence and using a synonym thesaurus.
All these methods do not perform well when faced with complex types of
plagiarism \citep[@][]{mozgovoy_automatic_2010} such as stealing ideas
or text translations. Modern approaches are based on methods for natural
language processing \citep{oberreuter_text_2013}, but they are still in
their infancy.

So far, several of the above-mentioned approaches to plagiarism
detection have been implemented in various types of software tools
varying from autonomous applications to web services. Typically,
applications are run locally and scanned for plagiarism within a given
corpus of documents. On the other hand, there are web services that
allow us to check for plagiarism among local corpuses and several
sources on the Internet.

The main role of the plagiarism detection software tool is detecting
similarities in program code, text or both. Some of the most commonly
used tools today for detecting plagiarisms in computer source code are
\textbf{Sherlock} \citep{joy_plagiarism_1999, mozgovoy_fast_2005},
\textbf{JPlag} \citep{prechelt_finding_2002} and \textbf{Moss}
\citep{schleimer_winnowing:_2003}. Their basic functionality is very
simple. Selected submissions are run through a similarity engine, which
provides pairwise results with potential plagiarisms. Modern software
for plagiarism detection in source code is based not only on methods for
string matching but also includes methods for searching lexical and
structural modifications in programming code
\citep{alzahrani_using_2012, alzahrani_understanding_2012, duric_source_2013, hein_scientific_2012, joy_plagiarism_1999, vrhovec_outsourcing_2015}.
On the other hand, \textbf{WCopyFind} \citep{balaguer_putting_2009},
\textbf{Ephorus} \citep{den_ouden_plagiarism:_2011} and
\textbf{TurnItIn}
\citep{buckley_evaluation_2013, marsh_turnitin.com_2004, rolfe_can_2011}
are tools for detecting plagiarisms in free text. They are used to find
the amount of text shared between two or more plain text documents on
the basis of fingerprinting
\citep{introna_sociomaterial_2011, mozgovoy_automatic_2010}.

The user-friendliness of all the above-mentioned applications varies
considerably. While some web services for detecting plagiarisms provide
an intuitive user interface (eg, Ephorus), the majority of tools require
a skilled user to operate them (eg, Moss) and they are not suitable for
use by an ordinary teacher with the average computer skills. However,
there is another consideration to be taken into account when using these
tools. The vast majority of them do not support the work of the
investigator through all four stages of the plagiarism detection process
(FSPDP). Current solutions are focused on the first and second stages
(mainly on the second stage) of the process, which means that the
investigator only gets a pairwise analysis (2nd stage) while the last
two stages must be performed manually.

Despite the many different approaches and tools available, researchers
\citep{mozgovoy_automatic_2010} have also pointed out that currently
available detection systems have several drawbacks which can be divided
into two main categories:

\begin{itemize}
\tightlist
\item
  issues concerning the user-friendliness of today's detection tools
  (implementation of the system) and
\item
  issues about the limitations of the existing technologies for
  plagiarism detection.
\end{itemize}

We also believe that the major drawback of current solutions is the
inability to support all of the four stages in the plagiarism detection
process. In fact, they can only be utilized in the first, and primarily
in the second stage of the FSPDP, while the latter two stages
(confirmation and investigation) have to be done manually by the
investigator (eg, teacher), thus extending the time needed for the
confirmation of plagiarized work and reducing overall efficiency of the
FSPDP.

Moreover, several studies and analyses of social networks were the
motivation behind our merging of information from social networks into
the plagiarism detection process to counter the drawbacks of current
approaches to plagiarism detection. Authors \citep{junco_too_2012} aimed
to identify relationship between Facebook use and academic performance.
The research confirmed a negative relationship between time spent on
Facebook and overall grade point average (GPA) achieved, as well as the
time spent preparing for class. There have been multiple research
studies conducted \citep{hew_students_2011, roblyer_findings_2010}
highlighting attitudes toward social networking sites and student and
faculty use of social networks. The results confirmed that students are
more likely to use social networking sites and are significantly more
open to the possibility of using similar technologies. Conclusions also
suggest that social networks have very little educational use, as they
are being used mainly to keep in touch with known individuals and
students tend to disclose more personal information about themselves on
social networking sites, hence exposing themselves to potential privacy
risks. Furthermore, authors \citep{subelj_expert_2011} successfully
employed social network information in detecting automobile insurance
fraud. The results provided some evidence that connectivity of users on
social networking sites can have a predictive value in determining
fraudulent activities like the detection of plagiarism.

\subsection{Problem and proposed
solution}\label{problem-and-proposed-solution}

The review of related approaches pointed out that the research studies
on plagiarism detection are focused on the first two initial stages -
collection (1st stage) and mainly on analysis (2nd stage). As we have
mentioned before, the majority of existing approaches conclude their
user support by providing information on the pairwise content similarity
of documents, and leaving the investigator to perform the confirmation
and investigation stages manually. In contrast, the proposed approach
puts the emphasis on an integrated solution where we try to focus on the
social aspects of possible plagiarists, by taking into account their
social network connections, activities, as well as information from the
Web. This provides improved support for plagiarism detection in
confirmation (3rd stage) and investigation (4th stage). We argue that
our approach facilitates plagiarism detection by providing the
investigator with better support in the latter two stages; therefore,
the confirmation or rejection of plagiarism in the third stage can be
more efficient, consequently making the process of plagiarism detection
also more efficient as a whole. The result of our approach represents
the reduced number of potentially plagiarized work that an investigator
has to examine manually, and the provision of new visualization
techniques that enable a semantically enriched view of relationships
among possible plagiarists. We also provide a tool to support the
process that can visualize more corpora with additional information.
This enables the investigators to have an overview of an author's
plagiarism in the context of their previous work and work related to
their colleagues, and not only by content similarity. The tool is
intended for use by teachers who want to exclude the possibility of
cheating among students.

In contrast to majority of the existing tools for plagiarism detection,
the provided tool has two important advantages:

\begin{itemize}
\tightlist
\item
  it is more user-friendly and as a consequence it can be used by any
  teacher with a common level of computer knowledge; and
\item
  it provides integrated support for an investigator in all four stages
  of FSPDP.
\end{itemize}

The tool is primarily designed for teachers or professors who teach
programming, enabling them to find plagiarism easily in source code
documents related to a particular assignment, or in a particular
teaching assignment when compared with all previous assignments.

\section{Social plagiarism detection
framework}\label{social-plagiarism-detection-framework}

\subsection{Description of proposed
system}\label{description-of-proposed-system}

\begin{figure}

{\centering \includegraphics[width=0.9\linewidth]{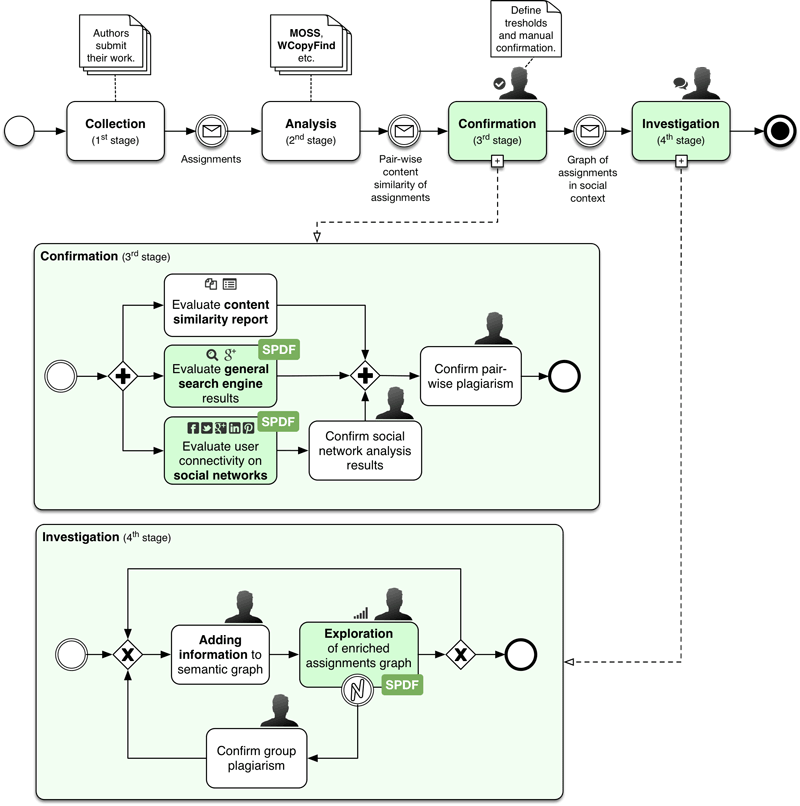}

}

\caption{Social Plagiarism Detection Framework (SPDF)}\label{fig:SPDF}
\end{figure}

With the introduction of the Social Plagiarism Detection Framework
(SPDF), we focus on the latter stages of the plagiarism detection
process, namely in confirmation (3rd stage) and investigation (4th
stage) as depicted in Figure \ref{fig:SPDF}.

The main contributions of our approach are as follows:

\begin{itemize}
\tightlist
\item
  integration of social network information and information from the Web
  that facilitates the plagiarism detection process; and
\item
  an advanced semantically enriched visualization (semantic graph,
  co-occurrence matrix) of information about authors and documents that
  enables the exploration of data in search of advanced patterns of
  plagiarism.
\end{itemize}

\begin{figure}

{\centering \includegraphics[width=1\linewidth]{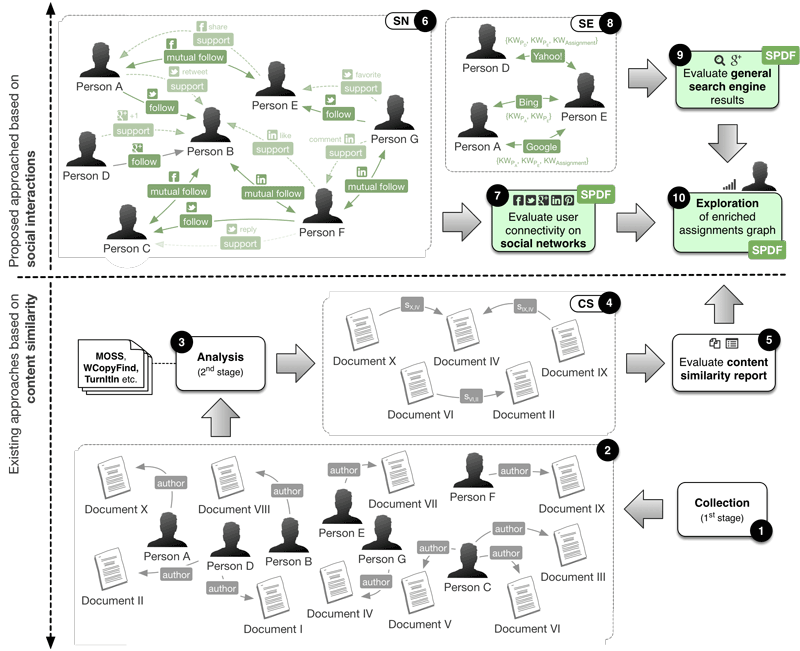}

}

\caption{Contribution of SPDF compared with existing approaches}\label{fig:SPDF-contribution}
\end{figure}

The additional steps of SPDF and advantages compared with existing
approaches based only on content similarity are depicted in Figure
\ref{fig:SPDF-contribution}. Steps 1 - 5 are generally performed in
content similarity matching in the plagiarism detection process, while
we introduce steps 6 - 10.

In the confirmation stage, the system evaluates the content similarity
report (Figure \ref{fig:SPDF-contribution}, step 5) provided in the
analysis stage (Figure \ref{fig:SPDF-contribution}, steps 3 - 4) and
performs an additional evaluation of general search engine results
(Figure \ref{fig:SPDF-contribution}, step 9) and connections between
authors on social networks (Figure \ref{fig:SPDF-contribution}, step 7).
In the case of ambiguity, the investigator is provided with an option to
review the social network analysis results. Based on all given
information in the context (content similarity and connections between
users on social networks), the investigator can confirm or reject
pairwise plagiarism. The main benefit of our approach is the improved
ranking of potential pairwise plagiarisms where social information as
well as information from the Web are taken into account, thereby
minimizing the effort required by the investigator in the confirmation
stage. We argue, and provide a comprehensive evaluation of our findings
in the fourth section, that the impact of social information is
statistically significant in the plagiarism detection process.

\subsection{Plagiarism detection framework
definition}\label{plagiarism-detection-framework-definition}

We can define \(P\) and \(D\) as nonempty sets of people and documents
respectively

\begin{equation}
P = \{ p \text{ | } p \text{ is a person} \}
\label{eq:P}
\end{equation}

\begin{equation}
D = \{d \text{ | } d \text{ is a document, written by } p \in P\}
\label{eq:D}
\end{equation}

where \(W\) is a set of pairs \(\langle p, d \rangle\) with \(p\) as an
author of a document \(d\)

\begin{equation}
W = \{ \langle p, d \rangle \text{ | } \exists p \in P \land d \in D : p \text{ is an author of } d \}
\label{eq:W}
\end{equation}

The set \(W\) is the result (Figure \ref{fig:SPDF-contribution}, step 2)
of the collection stage (Figure \ref{fig:SPDF-contribution}, step 1) and
is the direct input to the analysis stage (Figure
\ref{fig:SPDF-contribution}, step 3). For this purpose, we can further
define content similarity between documents

\begin{equation}
CS = \{ \langle d_i, d_j, s_{ij} \rangle \text{ | } \exists d_i, d_j \in D \land s_{ij} \in \left ( 0, 1 \right ] : s_{ij} \text{ is a share of content from } d_i \text{ in } d_j \}
\label{eq:CS}
\end{equation}

as a set of pairs of documents \(d_i\) and \(d_j\), with directed
content similarity \(s_{ij}\) between documents (Figure
\ref{fig:SPDF-contribution}, step 4).

By the integration of social network information in the plagiarism
detection process, we intro- duce two measures: \(SN\) (Figure
\ref{fig:SPDF-contribution}, step 6) and \(SE\) (Figure
\ref{fig:SPDF-contribution}, step 8).

\(SN\) represents the connections between users on social networks
considering pairwise actions between users. We define the following
types of actions
\(T(A) = \{ f^{\rightarrow }, f^{\leftrightarrow}, s^{\rightarrow} \}\)
as (directed) follow \(f^{\rightarrow}\), (undirected) mutual follow
\(f^{\leftrightarrow}\) and (directed) support \(s^{\rightarrow}\) (eg,
share, comment, reply, retweet, favorite, like, \(+1\) etc).

When considering various social networks, we classify them into two
distinct categories regarding user connections. The first group
considers links between users as directed (unilateral following of
users, eg, Twitter, Google+, etc), while the other group employs
undirected links (users mutually confirm following, eg, Facebook,
LinkedIn, etc). We can define the connections between users on social
networks

\begin{equation}
SN = \{ \langle p_i, p_j, A_{ij} \rangle \text{ | } \exists p_i, p_j \in P \land \exists a^{(ij)}_{k} \in A_{ij} : a \text{ is an action between } p_i \text{ and } p_j \}
\label{eq:SN}
\end{equation}

as a set of triples of people \(p_i\) and \(p_j\), with set of actions
\(A_{ij}\) between people \(p_i\) and \(p_j\). Actions are further
defined as set of triples
\(A_{ij} = \{ a_k^{(ij)} = \langle nt, act, w \rangle \}\), where \(nt\)
is the social network (eg, Facebook, Twitter, LinkedIn, Google+, etc),
\(act\) is activity (eg, follow, share, comment, like, etc) and \(w\) is
a user-defined weight of specific action.

When determining pairs \(\langle p_i, p_j \rangle\) of connected people,
a fuzzy search, implementing the Levenshtein distance algorithm, is
performed that requires further action by the investigator in the case
of ambiguity with multiple account and/or people matching.

We also introduce a set of related items from general search engine
\(SE\) between people \(p_i\) and \(p_j\) (Figure
\ref{fig:SPDF-contribution}, step 8) as

\begin{equation}
SE = \{ \langle p_i, p_j, KW_{ij}, n \rangle \text{ | } \exists p_i, p_j \in P, n \in \mathbb{N} : n \text{ is number of related items} \}
\label{eq:SE}
\end{equation}

where \(n\) is the number of relevant search results involving people
\(p_i\) and \(p_j\), using a set of keywords \(KW_{ij}\). This set of
keywords \(KW_{ij}\) between pairs \(\langle p_i, p_j \rangle\) of
connected people, is user defined per assignment as follows:

\begin{equation}
KW_{ij} = KW_{p_i} \cup KW_{p_j} \cup KW_{assignment}
\label{eq:KW}
\end{equation}

where \(KW_{p_i}\) and \(KW_{p_j}\) are keywords related to a person's
information (eg, name, surname, etc.) and \(KW_{assignment}\) is a set
of assignment-related keywords to narrow down the result set.

In the process of plagiarism detection, the goal is to define the set of
pairs of documents \(DP\), where plagiarism has been confirmed

\begin{equation}
DP = \{ \langle d_i, d_j \rangle \text{ | } \exists d_i, d_j \in D : d_i \text{ is a plagiat of } d_i \}
\label{eq:DP}
\end{equation}

When using existing approaches, the investigator, who performs the
plagiarism detection process, tries to identify elements of \(DP\),
while considering \(CS\) and some tacit knowledge \(TK\) by
investigation. The results of the confirmation and investigation stages
can be defined as a function \(check_{woSocio}\), performed by the
investigator

\begin{equation}
check_{woSocio} : CS \times TK \rightarrow DP
\label{eq:check-woSocio}
\end{equation}

The investigator evaluates the content similarity report consisting of
pairwise documents and decides on classifying the event as confirmed,
rejected or not-checked plagiarism.

\begin{figure}

{\centering \includegraphics[width=1\linewidth]{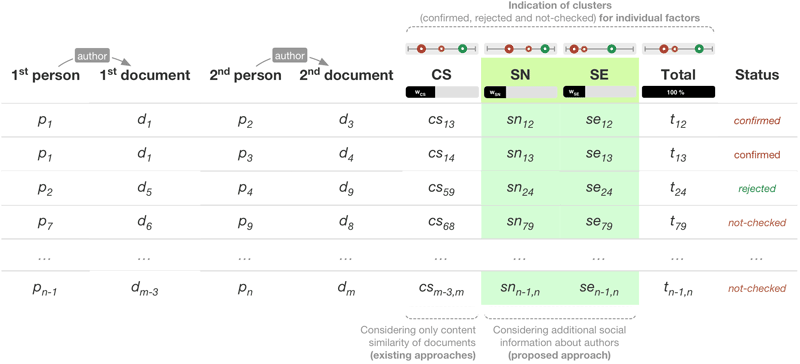}

}

\caption{Ranked table of pairwise possible plagiarism}\label{fig:ranked-table}
\end{figure}

When our proposed approach is utilized (Figure \ref{fig:ranked-table}),
the following function \(check_{wSocio}\) is defined

\begin{equation}
check_{wSocio} : CS \times TK \times SN \times SE \rightarrow DP
\label{eq:check-wSocio}
\end{equation}

that besides content similarity, also considers the social information
of document authors (Figure \ref{fig:SPDF-contribution}, steps 6 - 9).

Figure \ref{fig:ranked-table} depicts a table of pairwise potential
plagiarisms that the investigator has to examine in order to confirm or
reject them. The investigator is provided with the full information
about content similarity (\(CS\)), social network connectivity (\(SN\))
and general search engine matching (\(SE\)). The weighted average of the
aforementioned factors between people \(p_i\) and \(p_j\) is defined as

\begin{equation}
Total_{ij} = w_{CS} \cdot cs_{d(p_i) d(p_j)} + w_{SN} \cdot SN_{ij} + w_{SE} \cdot se_{ij}
\label{eq:total}
\end{equation}

where \(d(p_i)\) and \(d(p_j)\) are documents authored by people \(p_i\)
and \(p_j\) respectively.

The weights \(w_{CS}\), \(w_{SN}\) and \(w_{SE}\), where
\(\sum_i w_i = 1\), are user defined per assignment and allow
investigators to define the importance of individual plagiarism
detection factors (\(CS\), \(SN\) and \(SE\)).

To enable investigators to have an overview of factor values and their
distribution (Figure \ref{fig:ranked-table}), we introduce an indication
of clusters (\texttt{confirmed}, \texttt{rejected} and
\texttt{not-checked} possible plagiarism) for individual factors. It is
available as an interval with the minimum and maximum values for the
selected factor, where colors (\texttt{red}, \texttt{orange} and
\texttt{green} for \texttt{confirmed}, \texttt{not-checked} and
\texttt{rejected} potential plagiarism respectively) are used to depict
mean values for individual clusters.

Furthermore, we argue that the employment of \(check_{wSocio}\) is more
straightforward than \(check_{woSocio}\) and enables the investigator to
perform the confirmation stage more efficiently. This results in a
reduction of the total number of documents suspected to contain
plagiarism that the investigator has to manually review and confirm or
reject the occurrence of plagiarism. For evaluation purposes, the
supporting tool has been developed to test and compare the
aforementioned scenarios.

\subsection{Limitations}\label{limitations}

With the introduction of SPDF, the limitations of the proposed approach
also have to be considered. The major limitation of the approach is the
dependency on publicly available data from social networks. If we want
the approach to be as much efficient as possible, the authors (users)
should have publicly accessible profiles on several observable social
networks. In the case of Facebook, it is interesting that the
requirement is not so difficult to achieve, as studies about the user
identity presentation (user profile) on Facebook show that users are
willing to provide substantial amounts of personal data
\citep{gross_information_2005, wilson_review_2012}. Although the
awareness of privacy and security issues has increased over the past few
years, several studies have revealed that many users still have publicly
accessible profiles \citep{mazur_collecting_2010, mcknight_social_2011}.

The second limitation is that the approach is suitable for smaller
groups of authors (eg, in educational sectors or classes at University,
High School, etc) because it requires gathering user data from social
networks. Access to this kind of data is usually realized by using
application programming interfaces (APIs) which are generally limited in
terms of which and what data can be accessed and how often the data can
be retrieved \citep{rieder_studying_2013}. Facebook, for example, is
very restrictive in terms of which data can be accessed, and it also
determines the request frequency, making the possibility of analyzing
large set of users very impractical. Facebook, Twitter, LinkedIn and
others also reserve the right to close or modify these APIs, which
represent an additional limitation to the proposed approach.

The approach also raises some ethical questions as it anticipates the
automatic gathering of an author's data from social networks
\citep{zimmer_but_2010}. The authors of analyzed documents are not
active participants in the collection of their social data because they
could provide inaccurate data about social connections and their
activities on social networks. Therefore, they should be warned about
this approach used for plagiarism detection that uses automatic
inquiries about their profiles before their assignments are checked.
However, it is also important to point out that all the information
gathered about users is publicly available.

\subsection{Plagiarism detection assistant
tool}\label{plagiarism-detection-assistant-tool}

To support the proposed process, the plagiarism detection assistant
(PDA) tool was developed in which the following functionalities are
supported:

\begin{itemize}
\tightlist
\item
  creating and managing projects,
\item
  integration of existing plagiarism detection tools,
\item
  automatic acquisition of social network information and general search
  engine results,
\item
  confirming/rejecting assignments, and
\item
  advanced visualization.
\end{itemize}

The initial action in the process performed by an investigator is
creating a project and collecting the submissions. Then the following
steps include the preparation of data for the confirmation stage:

\begin{itemize}
\tightlist
\item
  performing content analysis by selecting the existing plagiarism
  detection tool, where pairwise content similarity report is retrieved;
  and
\item
  acquisition of social network information and general search engine
  results for investigated authors.
\end{itemize}

\begin{figure}

{\centering \includegraphics[width=0.8\linewidth]{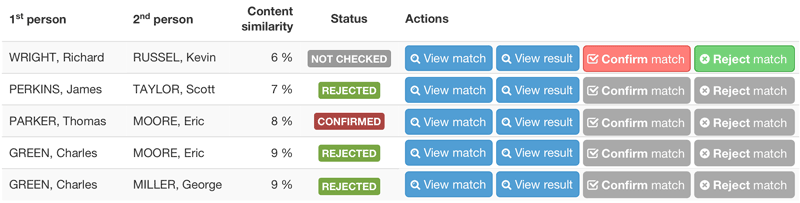}

}

\caption{Pairwise assignment view by content similarity}\label{fig:pairwise-view}
\end{figure}

After data are prepared, the investigator enters the confirmation stage
as Figure \ref{fig:pairwise-view} depicts (the names are pseudonyms).
The goal of this step is to assign one of the following status to the
pairwise assignments by two people who are being assessed for the
possibility of plagiarism:

\begin{itemize}
\tightlist
\item
  \texttt{not\ checked} - similarity between assignment has not been
  considered yet,
\item
  \texttt{rejected} - the assignments are not plagiarisms, and
\item
  \texttt{confirmed} --- the assignments contain plagiarized sections.
\end{itemize}

\begin{figure}

{\centering \includegraphics[width=0.8\linewidth]{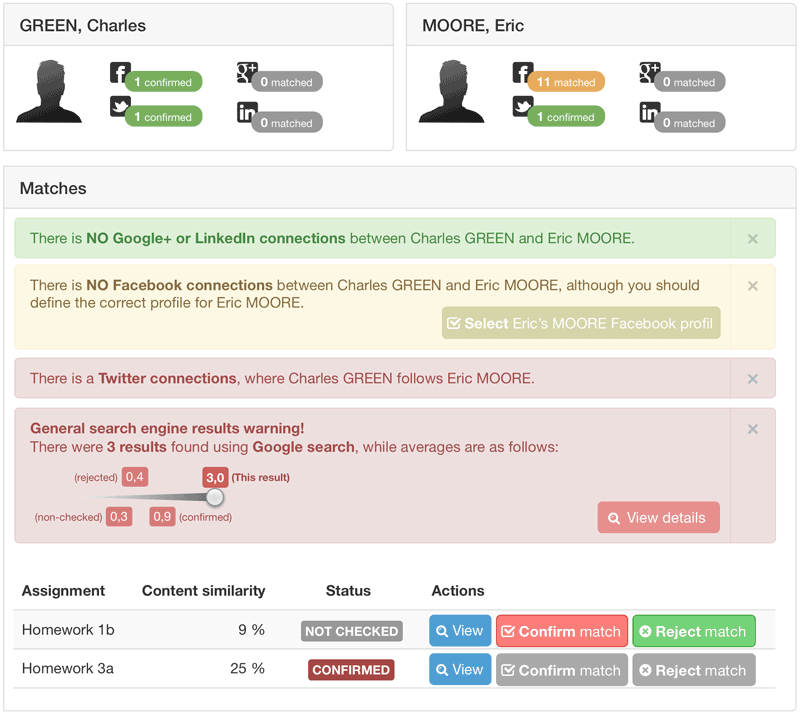}

}

\caption{Extensive report on matching, including social network information}\label{fig:extensive-report}
\end{figure}

When making the decision, the PDA tool assists the investigator by
providing an extensive report of matches found on assignments submitted
by different authors as depicted in Figure \ref{fig:extensive-report}.
There is a history of all assignments and their corresponding content
similarity enriched by the social network component. The colors used
depict the severity of the warnings. In this way, the tool automatically
ranks detected plagiarized sections depending on information obtained
from social networks and activities on the Web, but it does not confirm
suspicious assignments as plagiarism. We must emphasize that the
proposed approach is not intended to be used as a replacement for
conventional analyses of texts, but rather as a supporting tool for
increasing the efficiency of the confirmation stage. When the
investigator reviews all the provided information, they can make a
decision and confirm or reject plagiarism. By performing these steps,
the confirmation stage of the plagiarism detection process is concluded
(see Figure \ref{fig:SPDF}) and the investigation stage can begin.

\begin{figure}

{\centering \includegraphics[width=0.8\linewidth]{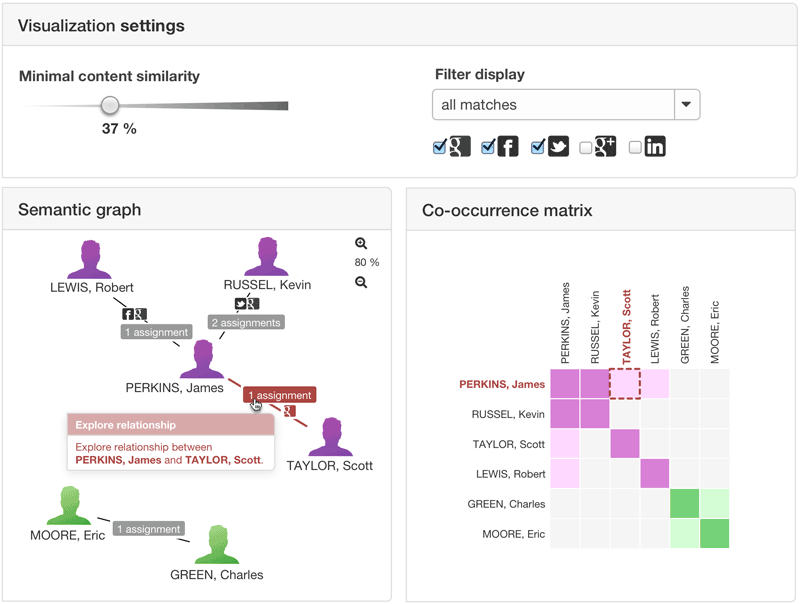}

}

\caption{Analytical assistance provided by the PDA tool}\label{fig:analytical-report}
\end{figure}

One of the views in the PDA tool within the investigation stage is
depicted in Figure \ref{fig:analytical-report} where the support for
advanced visualization is provided. The investigator can interactively
explore the semantic graph and co-occurrence matrix equipped with
information about the content similarity, connectivity on social
networks and general search engine results. The data from social
networks and the Web are collected by means of social network public
APIs and Web scraping of publicly available data about authors. As we
only do a pairwise analysis of data from a limited set of people, we do
not have any problems with processing resources. By visualizing the
context of the entire group under investigation (eg, class at
University), the investigator can carry out plagiarism detection by
exploring group plagiarism.

\section{Evaluation}\label{evaluation}

\subsection{Method}\label{method}

Our approach was evaluated on a case study of \(76\) students taking one
of the lectures in Computer Science at undergraduate level. Each student
had to submit five programming homework assignments during the semester
that were later checked for plagiarism. There were two experiments
performed with two groups of evaluators that followed different
approaches to the same dataset (\(76\) students submitted 5 assignments,
where \(22\) assignments were missing, so in total \(358\) assignments).
Both groups of evaluators had the common goal - to identify plagiarism
in the students' work. In the first approach, \(check_{woSocio}\)
evaluators employed MOSS \citep{schleimer_winnowing:_2003} and performed
a manual investigation on pairwise content similarity, while in the
second approach, \(check_{wSocio}\), our method with additional social
network analysis results was used. The information from social networks
employed in the second approach was extracted from the public profiles
of the students. In our case study, \(54\) students had publicly
available information on Facebook and \(43\) students were active on
Twitter. Students were informed about the use of all available public
information in the process of plagiarism detection throughout the
course.

The method used for the evaluation of the aforementioned approaches is a
generalized linear model with logistic regression where the link
function is defined as follows

\begin{equation}
g \left ( Y \right ) = log_e \left ( \frac{n}{1-n} \right ) = \beta_0 + \sum_{j = 1}^{p} \beta_j X_j
\label{eq:gY}
\end{equation}

The logistic regression is applied to a situation where the response
variable \(Y = cheat_{confirmed}\) is dichotomous \((0, 1)\). The model
assumes that \(Y\) follows a binomial distribution and it can be a fit
to a linear model \(g(Y)\). The conditional mean of \(Y\) is the
probability \(\pi = \mu_Y\) that cheat is confirmed, given a set of
\(X\) values. The odds that cheat is confirmed are \(\frac{n}{1-n}\) and
\(log \left ( \frac{n}{1-n} \right )\) is the log odds or \(logit\).

We have defined two models

\begin{equation}
check_{woSocio} : cheat_{confirmed} \sim match_{cs}
\label{eq:model-check-woSocio}
\end{equation}

\begin{equation}
check_{wSocio} : cheat_{confirmed} \sim match_{cs} + match_{fb} + match_{tw} + se_{hits}
\label{eq:model-check-wSocio}
\end{equation}

where \(check_{woSocio}\) is a nested model within \(check_{wSocio}\)
with the same response variable \(Y = cheat_{confirmed}\) and different
predictors \(X_{woSocio}\) and \(X_{wSocio}\), where
\(X_{woSocio} \subseteq X_{wSocio}\). The predictor variables are as
follows: \(cheat_{confirmed}\) is \(\{true, false\}\) factor with
information about confirmed plagiarism from investigator; \(match_{cs}\)
is content similarity \(s_{ij}\) between documents \(d_i\) and \(d_j\),
where \(\langle d_i, d_j, s_{ij} \rangle \in CS\); \(match_{fb}\) is a
\(\{ true, false \}\) factor, based on existence of
\(\left \langle p_i, p_j, \left \langle FB, follow, 1 \right \rangle \right \rangle \in SN\);
\(match_{tw}\) is a \(\{ true, false \}\) factor, based on existence of
\(\left \langle p_i, p_j, \left \langle TW, follow, 1 \right \rangle \right \rangle \in SN\)
and \(se_{hits}\) is a number of search engine results \(n\) between
people \(p_i\) and \(p_j\), where
\(\left \langle p_i, p_j, KW_{ij}, n \right \rangle \in SE\) and
\(KW_{ij}\) is a set of keywords including names and surnames of \(p_i\)
and \(p_j\) and subject title.

The employment of models \(check_{woSocio}\) and \(check_{wSocio}\) is
not intended to predict plagiarism, but rather for the ranking of
potential pairwise plagiarized sections that the investigator can review
and confirm in the latter stages of plagiarism detection.

To conclude our experiment, we performed a follow-through interviews
with all of the students where they defended their work and evaluators
determined if their submitted work was original. The evaluator's
decisions were then used for \(cheat_{confirmed}\) response variable to
evaluate both models.

\subsection{Results}\label{results}

When building a model checkwoSocio, the results show that the predictor
variable matchcs is significant
\(\left ( p \leq 9 \times 10^{-6} \right )\) in predicting the response
variable \(cheat_{confirmed}\).

The next step was to build another model \(check_{wSocio}\) with
integrated social network information, where the results of the second
model show that all predictor variables \(match_{cs}\)
\((p \leq 0,0025)\), \(match_{fb}\) \(\left (p \leq 0,0289 \right )\),
\(match_{tw}\) \(\left (p \leq 0,0904 \right )\) and \(se_{hits}\)
\(\left (p \leq 0,0432 \right )\) are significant in predicting the
response variable \(cheat_{confirmed}\).

Then we were able to compare both models, which consist of variables
that all have significant impact on the prediction. We performed ANOVA
with likelihood ratios test (LRT) on both models. The measure used for
comparison is deviance as a distance between two probabilistic models.
Deviance can be regarded as a measure of lack of fit between model and
data. Based on the results, we can conclude that the residual deviance
in the 1st model \(check_{woSocio}\) \(\left ( RD_1 = 48,932 \right )\)
is significantly higher \(\left ( p \leq 6 \times 10^{-8} \right )\)
than in the 2nd model \(check_{wSocio}\)
\(\left ( RD_2 = 12,479 \right )\). We can argue that the 1st model is a
poorer fit to the data and that the 2nd model performs better.

To confirm that the results are meaningful, we have performed the test
for overdispersion for both models that could lead to distort test
standard errors and inaccurate test of significance. We performed
fitting of the model twice - once with binomial family and second with
quasibinomial family and the results confirmed that overdispersion is
not a problem (the noncentral chi-squared test was not significant with
\(p_{woSocio} = 0,977\) and \(p_{wSocio} = 0,990\)). We also assessed
the model adequacy by checking for unusually high values in the hat
values, studentized residuals and Cook's D statistics. The results of
these tests also confirmed that the models are adequate.

\section{Discussion}\label{discussion}

As presented in the second section, the employment of advanced social
network analysis approaches has been already successfully implemented in
several business-oriented domains. In particular, our research was
motivated from fraud detection in car insurance industry where some
promising implementations can be found. The results of our approach
demonstrate that there also exists a predictive value in utilizing
social network information about authors of documents when detecting
plagiarism in student's work.

The main goal of SPDF is to make the overall process of plagiarism
detection (FSPDP) more efficient, which has several implications for
education in general by impacting all participants of FSPDP---students
and teachers. With more efficient support to plagiarism detection,
teachers can focus more on high-performing students, while
identification of ones that employ fraudulent actions is improved.
Still, the time invested by teachers also incorporates some additional
activities, such as disambiguation of students on social networks, when
system cannot determine the unique person from a list of available ones
(eg, people with the same name). But this invested time from teacher's
perspective is rewarded with reduced set of potential pairs of documents
to check for plagiarism. From the student's point of view, the
employment of their social network information in plagiarism detection
can have negative acceptance. This is a general concern, as people tend
to protect their own online privacy. But nonetheless, SPDF employs only
publicly available information from social networking sites and general
search engines. If there is no information available online about
students, then the proposed approach will not improve the plagiarism
detection process and we will have to utilize only content similarity
and perform all work in the 3rd and the 4th stages of FSPDP manually.

During the evaluation process of SPDF, there were several cases in our
study when investigators (eg, teachers) of plagiarism commended the
exploratory aspect of SPDF by supporting investigation and enabling them
the traversal of information about documents and authors in integrated
manner of a semantic graph and a co-occurrence matrix, thus making them
more efficient in evaluating the knowledge of students. Time invested to
plagiarism detection varied significantly betweentwogroupsof
investigatorsthatwereinvolvedinourstudy.Thefirstgroupthatemployed only
content similarity information had to review \(52\) pairs of assignments
identified as possible plagiarism. The other group using SPDF approach
identified only \(17\) pairs of possible plagiarism. The decreased
number of possible plagiarism to review and confirm was due to
availability of additional social network information and advanced
semantically enhanced visualization of results that enabled
investigators to identify clusters of students that collaborated and
therefore handling them in a group. The social information also often
provided investigators some intuition on who was the author (eg, general
search engine results indicated student is collaborating in an open
source project related to the subject of examination) and who was
copying. The enriched view on possible plagiarism also presented a
stronger evidence for investigators in the last stage of investigation
where students were invited to an interview and to defend their work.

Besides confirming that SPDF performs significantly better than a
content similarity approach in determining plagiarism of computer
programming assignments at University, we applied the approach in other
settings. One of the successful studies still in progress is in a High
School environment where students in a class for their mother tongue
language are required to write multiple essays about literary novels. As
there are several assignments per semester, instructors are overloaded
with work providing feedback to all students. The employment of SPDF was
therefore a welcome upgrade from manual examination and content
similarity evaluation process only. Instructors, with the help of SPDF,
more quickly and efficiently identify plagiarism and focus on providing
constructive feedback to students to support their progress.

\section{Conclusion and future work}\label{conclusion-and-future-work}

Plagiarism detection approaches mainly focus on the first two stages of
FSPDP. However, that is not sufficient to discover authors who perform
unethical acts relating to plagiarism, because we also have to deal with
false positive and false negative results from the analysis stage.

With the proposed approach and PDA software tool, we are able to support
the investigator's work effectively during the confirmation and
investigation stages. In the confirmation stage, we can efficiently
narrow the set of potential plagiarists from previous stages and in the
investigation stage we can visualize the relationships among potential
plagiarists with the additional semantic information. The evaluation of
two different models, in the selected case study, demonstrates that the
obtained results are significant. This provides evidence that the
inclusion of social network information about the authors of texts
assists the plagiarism detection process when compared with the approach
where the information from the social networks and the Web is not
employed during the manual decision-making process of confirmation and
investigation of plagiarism as performed by a human investigator. In
this place, it is also important to point out the restrictions of our
approach. The major limitation is its dependency on publicly available
data from social networks, inclination of users to provide public
personal information and possibility to access these data. There also
appear some ethical questions concerning automatic inquiries about user
profiles but it has to be emphasized that all data from social networks
and the Web that we utilized are publicly available.

Future research will focus on improving the framework by further
analysis of social network connections between authors under
investigation. The communication interactions will be analyzed by using
advanced methods of text analysis. We will also try to find the
correlation between the messages exchanged between authors and any
plagiarism in their submitted documents.

\end{document}